# Towards a scalable AI-driven framework for data-independent Cyber Threat Intelligence Information Extraction

1st Olga Sorokoletova
*Data & Artificial Intelligence*
CY4GATE S.p.A.
Rome, Italy
olga.sorokoletova@cy4gate.com

2nd Emanuele Antonioni
*Data & Artificial Intelligence*
CY4GATE S.p.A.
Rome, Italy
emanuele.antonioni@cy4gate.com

3rd Giordano Colò
*Data & Artificial Intelligence*
CY4GATE S.p.A.
Rome, Italy
giordano.colo@cy4gate.com

*Abstract*—Cyber Threat Intelligence (CTI) is critical for mitigating threats to organizations, governments, and institutions, yet the necessary data are often dispersed across diverse formats. AI-driven solutions for CTI Information Extraction (IE) typically depend on high-quality, annotated data, which are not always available. This paper introduces 0-CTI, a scalable AI-based framework designed for efficient CTI Information Extraction. Leveraging advanced Natural Language Processing (NLP) techniques, particularly Transformer-based architectures, the proposed system processes complete text sequences of CTI reports to extract a cyber ontology of named entities and their relationships.

Our contribution is the development of 0-CTI, the first modular framework for CTI Information Extraction that supports both supervised and zero-shot learning. Unlike existing state-of-the-art models that rely heavily on annotated datasets, our system enables fully dataless operation through zero-shot methods for both Entity and Relation Extraction, making it adaptable to various data availability scenarios. Additionally, our supervised Entity Extractor surpasses current state-of-the-art performance in cyber Entity Extraction, highlighting the dual strength of the framework in both low-resource and data-rich environments.

By aligning the system's outputs with the Structured Threat Information Expression (STIX) format, a standard for information exchange in the cybersecurity domain, 0-CTI standardizes extracted knowledge, enhancing communication and collaboration in cybersecurity operations.

## I. INTRODUCTION

Cyber Threat Intelligence (CTI) [1] is a fundamental discipline in cybersecurity that focuses on the collection, analysis, and interpretation of threat data concerning cyberattacks. This process involves collecting data from various sources, such as dark web forums, security incident databases, and network sensors, to build a comprehensive picture of potential threats. In this context, a key role is played by the Cyber Threat Intelligence analyst. The CTI analyst is a professional specializing in the collection, analysis, and interpretation of cyber threat intelligence. The CTI analyst identifies Indicators of Compromise (IOCs), analyzes attackers' Techniques, Tactics, and Procedures (TTPs), and provides strategic recommendations to improve security. Additionally, the CTI analyst collaborates with other security teams to implement preventive measures and respond to incidents, contributing to a secure digital environment. Despite several technological advances, the work of the CTI analyst still requires many manual knowledge extraction steps, such as data collection, verification, and analysis, which are laborious and time-consuming. Moreover, many companies, organizations, or public institutions lack the resources to hire a dedicated CTI analyst, exposing themselves to various risks. These risks include vulnerability to cyber attacks, loss of sensitive data, reputational damage, and potential breaches of data security regulations. The absence of a CTI expert can significantly undermine an organization's ability to defend itself against cyber threats. To address these problems, the industry has been investing heavily in the integration of Artificial Intelligence (AI) tools [2], [3]. AI enables the automation of many manual tasks, such as network monitoring and data analysis, increasing the efficiency and speed of the intelligence process. For example, Machine Learning (ML) algorithms can identify patterns and anomalies in data, signaling imminent threats for faster and more accurate responses. Additionally, AI helps predict future threats by analyzing historical and behavioral trends of attackers.

In line with the intention of integrating AI techniques effectively within the CTI analyst's workflow, this paper introduces 0-CTI, an innovative system designed to address various challenges in the field of Cyber Threat Intelligence. The system employs a fully Machine Learning-based approach to extract knowledge in a structured format from unstructured natural language texts. 0-CTI was developed to significantly accelerate the work of CTI analysts by automating numerous laborious steps in the data extraction and analysis process. Traditionally, analysts spend considerable time manually collecting, verifying, and interpreting threat information. With 0-CTI, these tasks can be quickly and accurately performed, enabling analysts to focus on more strategic and decision-making responsibilities. Another major advantage of 0-CTI is its ability to support organizations lacking a dedicated

CTI analyst. Automating knowledge extraction steps allows these organizations to mitigate risks associated with the absence of specialized expertise. Furthermore, 0-CTI generates knowledge in a STIX-compliant format, which is a widely recognized standard in the community. This facilitates universal information sharing, improving collaboration between different organizations and strengthening the collective ability to defend against cyber threats. Thanks to the compliance with STIX, extracted information can be easily integrated into existing systems, enhancing the effectiveness of cybersecurity operations.

0-CTI represents a remarkable technological advancement in CTI, harnessing the potential of Transformer neural networks to achieve superior performance in state-of-the-art Entity and Relation Extraction. By utilizing these advanced networks, 0-CTI can accurately recognize and analyze a wide range of entities and their interconnections, overcoming limitations of existing solutions, such as the limited number of classes and strong data dependence. In addition to the traditional NER system based on prior dataset training, our system proposes a zero-shot variant of the Entity Extraction model. This variant employs single-level taxonomies of STIX-compliant classes combined with zero-shot Named Entity Recognition (NER) systems like GLiNER [4]. 0-CTI can recognize and categorize new or unknown entities without requiring a preliminary training phase, making the integration and update process much faster and more flexible. Additionally, 0-CTI employs a novel algorithm based on cross-encoder neural networks to extract relationships between entities in zero-shot mode. This approach allows the system to identify connections between different entities without specific training, further expanding its capabilities in analyzing and correlating information. The final result is a knowledge graph that embeds all relevant information expressed by the original plain text. With these features, 0-CTI can seamlessly scale to diverse applications, independent of an organization's capacity to deploy specialists for data preparation and validation. This means that even organizations without dedicated CTI resources can benefit from an advanced, automated system for threat extraction and analysis, significantly improving their cybersecurity awareness.

The paper is organized as follows: Section II formulates the task of Information Extraction (IE) in compliance with Structured Threat Information eXpression (STIX) standard. Section III compares the system to current AI endeavors in CTI. Sections IV and V detail the architecture, methodologies, and experimental evaluation of 0-CTI. Finally, Section VI concludes with the system's contributions and future developments.

## II. INFORMATION EXTRACTION OF STIX OBJECTS

Structured Threat Information eXpression (STIX) was introduced as a language for the systematic and machine-readable representation of CTI data in a publication [5] authored by the MITRE Corporation, a research organization distinguished for its noteworthy contributions to the field of Cybersecurity.

Acknowledged as a widely adopted standard within the CTI community, the STIX has established itself by proclaiming a systematic classification covering all aspects of suspicion, compromise, and attribution. With objects and descriptive relationships, this systematic classification serves the purpose of enabling a consistent data-sharing among organizations. In a more formal context, STIX can be characterized as a schema that defines a taxonomy of Cyber Threat Intelligence, comprising six distinct classes, three of which are the primary focus of our attention: *STIX Domain Objects (SDOs)* that are higher-level intelligence objects, representing behaviors and constructs that threat analysts typically engage with to comprehend the threat landscape; STIX Cyber-observable Objects (SCOs) that encapsulate *Indicators of Compromise (IOCs)*, a perennial concern in Cybersecurity; and *STIX Relationship Objects (SROs)* that connect SDOs together, SCOs together, and SDOs with SCOs.

Given that our work addresses the Information Extraction problem, it is crucial to align this formulation with STIX terminology to standardize the system's output, facilitating efficient information sharing within the cybersecurity community.

The task of Information Extraction from textual data, as outlined by Jakub Piskorski and Roman Yangarber in [6], involves automating the identification and extraction of structured information or knowledge from unstructured text. It encompasses the identification and extraction of specific entities, relationships, and events mentioned in the text, transforming raw data into a structured format that is easily processable and analyzable. In our case, the holistic task breaks down into two subtasks, which can be tackled either concurrently or sequentially: *Entity Extraction (EE)* and *Relation Extraction (RE)*. Within the framework of STIX, entities for Entity Extraction are represented by a subset of SDOs and SCOs, while relationships in Relation Extraction are defined by SROs.

## III. RELATED WORKS

The necessity to automate Cyber Threat Intelligence data processing, emphasizing Entity and Relation Extraction in compliance with community standards, has driven researchers to embrace Information Extraction methodologies in Cybersecurity. Despite significant efforts (Gasmi et al. 2019 [7]; Legoy et al. 2020 [8]), many current methods struggle with the volume of extracted entities and relations and do not fully adhere to the STIX taxonomy. However, the adoption of Artificial Intelligence (AI), particularly Transformer models, is enhancing the efficiency of these processes, facilitating more robust and automated CTI analysis.

Weerawardhana et al. (2015) [9] compared a Machine Learning-based approach and a Part-of-Speech tagging method for Information Extraction, focusing on vulnerability databases. Li et al. (2019) [10] introduced a self-attention-based Neural Network for Cybersecurity Named Entity Recognition (NER), integrating features from words and characters using Convolutional Neural Networks (CNN) and a self-attention mechanism based on Bidirectional Long Short-Term

Memory (BiLSTM) and Conditional Random Field (CRF) models.

Ranade et al. (2021) [11] developed CyBERT, a domain-specific BERT model fine-tuned on a Cybersecurity corpus, which was further improved by retraining on a designated corpus tailored to STIX entities. This model outperformed others in the Massive Text Embedding Benchmark (MTEB) Leaderboard (Muennighoff et al., 2022a [12]).

Wang et al. (2022) [13] presented an Entity Recognition model that uses BERT for dynamic word vectors and BiLSTM-CRF for word sequence encoding, refined with CTI-specific knowledge engineering. Alam et al. (2022) [14] introduced CyNER, an open-source Python library for Cybersecurity NER, utilizing Transformer-based models and heuristics for extracting Indicators of Compromise (IOCs).

Zhou et al. (2022) [15] developed CTI View, an automated system for extracting and analyzing unstructured CTI associated with Advanced Persistent Threats (APTs), employing a Text Extraction framework and a BERT-BiLSTM-CRF model enhanced with a Gated Recurrent Unit (GRU) layer. Zhou et al. (2023) [16] later presented CDTier, a CTI dataset emphasizing threat Entity and Relation Extraction, improving model accuracy in extracting knowledge objects and relationships.

Marchiori et al. (2023) [17] introduced STIXnet, a modular solution for extracting STIX Objects from CTI Reports. STIXnet, aligned with the STIX taxonomy, represents the state-of-the-art in Information Extraction within the domain, but our system demonstrates superior performance in Entity and Relation Extraction compared to STIXnet. Unlike STIXnet's rule-based approach, our 0-CTI leverages Transformers for Entity Extraction and a novel cross-encoder model for Relation Extraction, employing a dataless training approach.

Our zero-shot Named Entity Recognition (NER) model, based on GLiNER [4], addresses shortcomings identified in previous studies [18], [19], [20], [21], [22], [23], and [24], such as slower processing speeds, large parameter sizes, and limited capability to predict multiple entity types concurrently.

In conclusion, 0-CTI stands out as a unique system capable of simultaneous Entity Extraction and Relation Extraction, leveraging advanced AI-based methodologies to identify a comprehensive range of STIX-compliant entities and relationships, operating in a dataless manner, and advancing the state-of-the-art in CTI Entity Extraction.

## IV. 0-CTI SYSTEM

The 0-CTI, a modular Transformers-based system, serves as a "lens" applied to raw, unstructured collections of Cyber Threat Intelligence, zooming in on STIX Domain Objects, Indicators of Compromise (IOCs), and their relationships expressed as STIX Relationship Objects. The overall architecture of the system is delineated in the block scheme depicted in Fig. 1, with a detailed explanation of each component provided in this section.

### A. Dataset and Text Processing

In the initial phase, the *Text Processing* module ingests CTI documents and processes the raw textual content within, producing a dataset containing sanitized, tokenized, and labeled text chunks. This dataset is provided as the output of this module and the input for the subsequent modules.

The system accepts CTI documents in English in various formats, including PDF, DOCX, or HTML. The extraction process identifies 9 entity classes corresponding to 9 STIX Domain Objects[1], and 21 relation classes corresponding to 21 STIX Relationship Objects. The dataset used for training the supervised NER model contained annotations for entities, with statistics presented in Table I, but did not include annotations for relations, which are: *ATTRIBUTED_TO, AUTHORED_BY, BEACONS_TO, COMMUNICATES_WITH, COMPROMISES, CONSISTS_OF, CONTROLS, DELIVERS, DOWNLOADS, DROPS, EXFILTRATE_TO, EXPLOITS, HAS, HOSTS, IMPERSONATES, INDICATES, LOCATED_AT, ORIGINATES_FROM, OWNS, TARGETS, USES*.

The Identity SDO is categorized to distinguish between persons and organizations, while the Indicator SDO is subdivided to encompass 23 types of IOCs/SCOs within the dataset, including domain names, email addresses, file hashes, IP addresses, URLs, and registry key paths, among others (for a comprehensive list and dataset statistics on their occurrences, refer to Table II).

The responsibility for discovering these indicators lies with the IOC-finder. Concurrently, the input presented to the ML-driven core of Entity Extraction undergoes a masking process using strings associated with the respective IOC types to prevent the learning process from being disrupted by erratic tokenization.

The aggregation of data from diverse sources inherently gives rise to noise, necessitating comprehensive preprocessing prior to initiating the training phase. Our text processing workflow is structured into three stages: standardization and artifact removal, chunking and recalibration, and CoNLL-formatting.

---

[1]The STIX taxonomy comprises 18 SDOs; however, certain ones are not pertinent to the NER task and instead serve as workaround instruments (for instance, the Grouping SDO is used to combine other objects).

TABLE I
THE DISTRIBUTION OF THE LABELED ENTITIES IN THE DATASET.

| Entity Type | Count |
| --- | --- |
| ATTACK_PATTERN | 2993 |
| CAMPAIGN | 553 |
| IDENTITY_ORGANIZATION | 2633 |
| IDENTITY_PERSON | 551 |
| LOCATION | 6782 |
| MALWARE | 10902 |
| THREAT_ACTOR | 6228 |
| TOOL | 2529 |
| VULNERABILITY | 786 |
| **TOTAL** | **33957** |

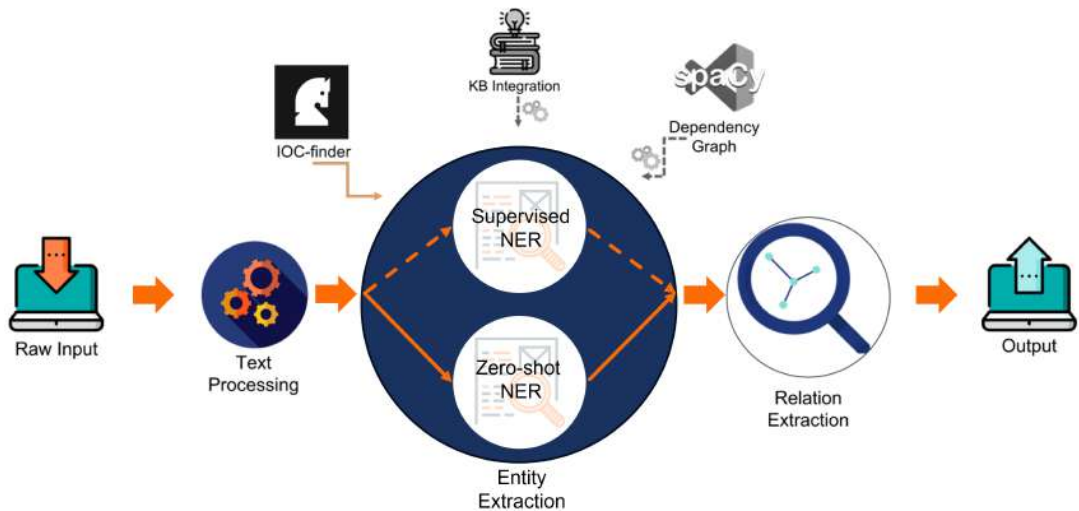

Fig. 1. 0-CTI pipeline. Orange bold connectors denote the mandatory integration of the component in the system and a completely dataless flow, orange dashed connectors denote integration of the supervised NER, which requires annotations for entities in the dataset but provides state-of-the-art performance for Entity Extraction, whereas gray connectors indicate optional extra blocks that can be integrated into the system, if desired by application.

TABLE II
THE DISTRIBUTION OF THE LABELED IOCs IN THE DATASET.

| IOC Type | Count |
| --- | --- |
| INDICATOR_ATTACK_TACTICS_ENTERPRISE | 49 |
| INDICATOR_ATTACK_TACTICS_MOBILE | 1 |
| INDICATOR_ATTACK_TECHNIQUES_ENTERPRISE | 789 |
| INDICATOR_ATTACK_TECHNIQUES_MOBILE | 1 |
| INDICATOR_BITCOIN_ADDRESSES | 484 |
| INDICATOR_CVES | 700 |
| INDICATOR_DOMAINS | 8716 |
| INDICATOR_EMAIL_ADDRESSES | 728 |
| INDICATOR_FILE_PATHS | 800 |
| INDICATOR_IMPHASHES | 5 |
| INDICATOR_IPV4S | 2049 |
| INDICATOR_IPV4_CIDRS | 27 |
| INDICATOR_MAC_ADDRESSES | 4 |
| INDICATOR_MD5S | 3816 |
| INDICATOR_MONERO_ADDRESSES | 5 |
| INDICATOR_REGISTRY_KEY_PATHS | 260 |
| INDICATOR_SHA1S | 1559 |
| INDICATOR_SHA256S | 6463 |
| INDICATOR_SHA512S | 4 |
| INDICATOR_SSDEEPS | 2 |
| INDICATOR_TLP_LABELS | 9 |
| INDICATOR_URLS | 1609 |
| INDICATOR_USER_AGENTS | 23 |
| TOTAL | 28103 |

- The initial processing stage involves the elimination of undesired elements such as log file names, redundant separators, unsupported characters, and sanification. Furthermore, our approach encompasses handling artifacts such as IOC tables, images, and code snippets that may appear in the reports.
- Following the completion of raw text processing, the next step involves segmentation into chunks using a recursive chunkizer methodology. This approach is designed to capture distant relationships between entities dispersed across different sentences, acknowledging that many reports contain relations that span beyond the typical chunk boundaries. Unlike non-recursive chunkizers that do not allow overlaps and may overlook factual relationships present in the report, our approach ensures comprehensive coverage for the Relation Extraction (RE) module analysis. Then, the recalibration process is executed to adjust the positions of entities in annotations.
- After the preceding steps, the Text Processing module finalizes its operation, generating an output dataset that encompasses the processed data formatted according to the specifications outlined in the CoNLL-2003 dataset.

### B. Entity Extraction with Supervised Named Entity Recognition and IOC-finder

The subsequent module in Fig. 1 is dedicated to *Entity Extraction* and consists of two primary submodules: the core extractor and the IOC-finder. The core extractor is embodied by the currently leading Transformer model for Token Classification, fine-tuned on an annotated dataset, and the zero-shot NER model, GLiNER. The IOC-finder is regarded as a mandatory sub-component. The modularity of the framework is highlighted by the gray connectors in the block scheme, which indicate the potential to integrate additional sub-components to enhance the core extractor. Examples of such augmentations include an interactive Knowledge Base (KB) to facilitate continuous system improvement, or a SpaCy POS Tagger and Dependency Parser to improve the system's ability to discover new entities and relationships based on observable linguistic patterns, without relying solely on AI.

The IOC-finder serves the purpose of identifying Indicators of Compromise (IOCs) submodule. The submodule was implemented using the utilities of Floyd Hightower's open-source project/library: IOC Finder. It operates in precedence to the core submodule and supplements its Machine Learning approach with the traditional NLP RegEx technique. The rationale behind it lies in the structured nature of IOCs, enabling their efficient detection through pattern matching. Typically, IOCs are presented in a list or table at the end of a CTI report and do not constitute integral parts of sentences, lacking contextualization cues. Moreover, certain IOCs, such as hashes, exhibit a non-linguistic high-entropy structure. When tokenized, these can yield a plethora of extraneous tokens, potentially impeding the effectiveness of fine-tuning the Transformer model and negatively impacting its training process.

To implement supervised NER in the Entity Extraction core, we have employed a model from the family of BAAI General Embedding (BGE) models[2]. Introduced by Xiao et al. in 2023 [25], BGE models represent Chinese Text Embedding Models (C-TEM) and include a versatile range of BERT-based embedding models across various scales: large (326M parameters), base (102M parameters), and small (24M parameters). In addition to general Chinese embeddings, the authors have also provided data and models specifically designed for English text embeddings. Acknowledged for their efficacy, BERT-based models have emerged as preeminent performers in the domain of Named Entity Recognition[3], with BGE currently representing the pinnacle in achieving an optimal balance between operational efficiency and representation quality. Hence, for our task, we opt for the `BAAI/bge-base-en-v1.5`[4] variant of the model to serve as our backbone Transformer model.

The computed confidence score of the BGE-based model after fine-tuning is subsequently utilized for merging the extraction results with those of the zero-shot extraction in applications where annotated data are available. This process includes a preliminary step wherein false positives are discarded through a thresholding operation applied to both NER outputs.

Upon the completion of execution by all Entity Extraction submodules, their respective outputs are integrated both intermodularly and with the outcomes of the Text Processing module. This integration yields a composite input consisting of processed texts and the extracted entities, including their relative positions within the texts. This composite input serves as the input for the subsequent *Relation Extraction* module. It is important to note that the sequential architecture for Entity Extraction and Relation Extraction modules, while potentially degrading the performance of the latter when the former fails to recognize entities effectively, was selected over a conjoint architecture to uphold the modularity of the system.

### C. Zero-Shot Components

Zero-shot learning empowers models to predict classes they have never encountered during training, revolutionizing tasks such as Named Entity Recognition. Traditional NER models rely heavily on annotated datasets. Zero-shot NER overcomes this limitation by utilizing models pretrained on extensive, diverse datasets and applying them to unseen entity types.

Another significant development is zero-shot classification through cross-encoders. Cross-encoders, which process pairs of input sequences together, excel at capturing intricate relationships between text segments. By employing cross-encoders for zero-shot classification, models can assess and classify texts based on contextual cues and relational patterns learned from vast datasets.

This chapter explores the implementation of zero-shot techniques in 0-CTI. We introduce an innovative system that combines GLiNER (Generalist and Lightweight Model for Named Entity Recognition) with a flat taxonomy system to achieve zero-shot Entity Extraction. Additionally, by employing cross-encoders, the substitution algorithm can infer and extract relations based on the contextual interplay between entities, even in the absence of explicit training data. This advancement opens new avenues for understanding complex inter-entity dynamics in diverse textual contexts.

*1) Zero-Shot Named Entity Recognition:* Our system performs zero-shot NER by combining GLiNER [4] with a specific paradigm of class substitution based on a flat taxonomy. Each entity class that the system has to recognize is divided into several child categories. If a token is assigned to one of these child labels, it is automatically assigned to the parent class. For example, the MALWARE class is mapped to subclasses such as Malicious Software, Trojan, Ransomware, and more.

We have chosen GLiNER as the only available zero-shot Entity Extractor that does not rely on resource-intensive Large Language Models, which can be difficult to deploy. This selection reinforces the core idea of our framework: to provide a ready-to-use solution accessible for a wide range of applications, even in resource-limited conditions.

At its core, GLiNER leverages pretrained language models like BERT or GPT to comprehend and infer relationships between words and entities within a text, independent of prior exposure to specific entity types. GLiNER operates through a structured methodology for entity recognition. When given a text, it initially encodes the input using a pretrained model, capturing contextual embeddings of the words. The next step involves applying a zero-shot inference mechanism. GLiNER interprets Entity Extraction as a Natural Language Inference (NLI) problem wherein the model is tasked with determining whether a given hypothesis (the presence of a specific entity type) is supported by the premise (the input text).

Combining the capabilities of GLiNER approach with a taxonomy of classes derived from our mapping system, we successfully integrated a zero-shot Named Entity Recognition alternative into the Entity Extraction core of 0-CTI. This feature is particularly advantageous for users seeking to extend the system's capability to recognize additional classes beyond those covered by our supervised NER model, eliminating the necessity of acquiring and annotating new data.

---

[2]Beijing Academy of Artificial Intelligence
[3]Token Classification Models
[4]BAAI/bge-base-en-v1.5

*2) Relation Extraction using Cross-Encoders:* Cross-encoders [26] are powerful NLP model, distinguished for their proficiency in assessing the semantic relationships between pairs of text sequences. Unlike bi-encoders, which encode each sequence independently, cross-encoders process both sequences together, enabling them to capture intricate dependencies and context. Thanks to this feature, cross-encoders demonstrate particular effectiveness in tasks such as assessing the implication of one sentence by another, which is crucial for diverse AI applications, including Relation Extraction. A cross-encoder takes a pair of sentences as input and assesses whether the meaning of the second sentence is entailed by the first.

For the *Relation Extraction* module in Fig. 1, we propose an algorithm leveraging cross-encoders to extract relations from text. This algorithm integrates the outputs of the Entity Extraction module and adheres to the STIX standard for defining relations.

- **Entity Extraction**: First, IOCs and entities within the text are extracted using the IOC-finder and NER systems: the zero-shot NER model, the supervised NER model, or a combination of both.
- **Potential Relations Creation**: Once entities are identified, the algorithm proceeds to generate potential relations by associating entities with their corresponding entity types, referencing the table of potential SROs for given pair of SDOs, which is provided in the STIX documentation[5]. For each pair of entities, the algorithm formulates a set of candidate sentences following the pattern: $<Entity1><Relation><Entity2>$. For instance, if the text mentions "APT1" and "Microsoft", a possible relation could be expressed as "APT1 targets Microsoft."
- **Relation Evaluation with Cross-Encoders**: The cross-encoder takes both the original text and the candidate relation sentence as input and calculates the likelihood that the candidate relation is true based on the context of the original text.
- **Threshold-Based Validation**: The algorithm assigns a truth value to each candidate relation given the score provided by the cross-encoder. Relations that surpass a predefined threshold are considered valid. This thresholding operation is crucial in ensuring that only the most probable relations are retained, thereby enhancing the accuracy of the extraction process.
- **Relation Disambiguation**: In cases of ambiguity, such as when the same relation occurs between two entities in both directions (e.g., $<Entity1><Relation><Entity2>$ and $<Entity2><Relation><Entity1>$), the algorithm disambiguates by comparing the cross-encoder scores. The relation with the higher score is retained, ensuring selection of the most contextually accurate relation.

[5]https://oasis-open.github.io/cti-documentation/stix/intro

By following this approach, the algorithm performs precise and contextually aware Relation Extraction, generating a cyber graph as exemplified in Fig. 2. Through methodical evaluation and disambiguation procedures, it ensures robust performance across diverse textual data.

## V. EXPERIMENTAL EVALUATION

At the time of writing this paper, our experimental investigations into the 0-CTI framework are ongoing, focusing on collecting a dataset annotated for cyber relations to enable a comprehensive assessment of the Relation Extraction module and, consequently, the overall system performance. In the meantime, we present a rigorous quantitative performance evaluation of the supervised Named Entity Recognition model within the Entity Extraction core. This is complemented by a qualitative assessment of our zero-shot components, which leverage advanced Large Language Models (LLMs). Together, these evaluations highlight the system's potential impact in the field of Cyber Threat Intelligence.

### A. Performance Evaluation of the Supervised Entity Extractor

To evaluate the performance of the fine-tuned Named Entity Recognition model, we conducted two experiments. The first experiment compared the effectiveness of two Transformer models and a Word2Vec embedding with an LSTM on dataset that we collected from OpenCTI sources. In the second experiment, we benchmarked the best-performing model from the first experiment against STIXnet [17].

In both experiments, we began by identifying Indicators of Compromise through the IOC-finder submodule, replacing them in the text with placeholders. Given the IOC-finder's accuracy, approaching $100\%$, we excluded its results from the evaluation to concentrate on the submodules that require contextual awareness.

*1) Models Comparison on Our Dataset:* In this initial experiment, we conducted an 18-class classification task, employing 9 classes with BIO-tagging. The dataset was divided into a training set comprising 3500 text chunks and a test set with 880 chunks. Since some chunks do not contain entities, we expect empty outputs for these instances and aim to minimize False Positives.

Both Transformer models were fine-tuned for 12 epochs using `Cross-Entropy` loss, the `AdamW` optimizer, a batch size of 4, and a learning rate of $2e^{-5}$. The first network, 0-$CTI_{BGE}$, utilized the `BGE-en-v1.5` embeddings [25], known for their strong general-purpose natural language capabilities. The second model, $0\text{-}CTI_{CyBERT}$, employed embeddings specifically trained for Cybersecurity tasks [11].

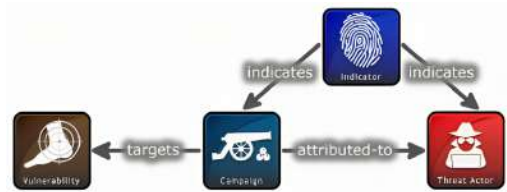

Fig. 2. STIX cyber graph example. Image credit: Introduction to STIX.

TABLE III
COMPARISON OF THE PERFORMANCES OF THE 0-CTI BGE CORE, 0-CTI CYBERT CORE, AND WORD2VEC + LSTM CORE FOR SUPERVISED NER IN ENTITY EXTRACTION.

| Entities | | Train F1-score | | | Test F1-score | | |
|---|---|---|---|---|---|---|---|
| Type | Count | $0\text{-}CTI_{BGE}$ | $0\text{-}CTI_{CyBERT}$ | LSTM | $0\text{-}CTI_{BGE}$ | $0\text{-}CTI_{CyBERT}$ | LSTM |
| ATTACK_PATTERN | 2993 | 0.95 | **0.97** | 0.96 | 0.64 | **0.67** | 0.42 |
| CAMPAIGN | 553 | **0.98** | **0.98** | 0.94 | **0.73** | 0.72 | 0.46 |
| IDENTITY_ORG | 2633 | 0.96 | **0.99** | 0.91 | **0.61** | **0.61** | 0.27 |
| IDENTITY_PER | 551 | 0.99 | **0.99** | 0.95 | **0.75** | 0.67 | 0.26 |
| LOCATION | 6782 | 0.99 | **1.00** | 0.98 | **0.89** | **0.89** | 0.72 |
| MALWARE | 10902 | **0.99** | 0.99 | 0.94 | **0.91** | **0.91** | 0.35 |
| THREAT_ACTOR | 6228 | 0.99 | **1.00** | 0.97 | **0.92** | **0.92** | 0.47 |
| TOOL | 2529 | 0.98 | **0.99** | 0.93 | **0.66** | **0.66** | 0.23 |
| VULNERABILITY | 786 | 0.98 | **0.99** | 0.97 | 0.90 | **0.92** | 0.33 |
| Weighted average | 33957 | 0.98 | **0.99** | 0.95 | **0.85** | **0.85** | 0.40 |

TABLE IV
COMPARISON OF THE PERFORMANCES OF THE 0-CTI CYBERT CORE AND STIXNET IN ENTITY EXTRACTION FOR THE MOST FREQUENT ENTITY TYPES IN THE DATASET USED IN [17].

| Entities | Evaluation F1-score | |
|---|---|---|
| | $0\text{-}CTI_{CyBERT}$ | $STIXnet$ |
| ATTACK_PATTERN | **0.99** | 0.77 |
| CAMPAIGN | **0.94** | 0.41 |
| IDENTITY | **0.99** | 0.79 |
| LOCATION | **0.99** | 0.94 |
| MALWARE | **0.98** | 0.84 |
| INTRUSION_SET | **0.99** | 0.94 |
| TOOL | **0.98** | 0.70 |
| Overall | **0.98** | 0.77 |

The LSTM was trained over 200 epochs using `Cross-Entropy` loss, the `Adam` optimizer, a batch size of 64, and a learning rate of $1e^{-3}$. This model consisted of 2 LSTM layers, 2 Dense Hidden Layers, each with 1024 neurons, culminating in an Output layer. We used the `GoogleNews-vectors-negative300` as pretrained Word2Vec embeddings.

The results of the evaluation are summarized in Table III, with the F1 score as the evaluation metric. Both Transformer-based models demonstrated comparable performance, underscoring their robustness and adaptability. The sub-module achieved a maximum test F1 score of 0.85, with the $0\text{-}CTI_{CyBERT}$ model emerging as the top performer. Conversely, the LSTM model exhibited significantly lower performance, attributed to the limitations of its pretrained embedding system in capturing contextually relevant information.

Further analysis revealed that the Transformer submodules experienced a loss of generalization across specific classes, such as ATTACK_PATTERN, CAMPAIGN, IDENTITY_PERSON, IDENTITY_ORGANIZATION, and TOOL. The imbalance in dataset likely contributed to the challenges observed with IDENTITY_PERSON and CAMPAIGN. Additionally, the nuanced nature of campaign concepts may complicate recognition, even for human readers. Notably, entities such as ATTACK_PATTERN and IDENTITY_ORGANIZATION often span multiple words, making their extraction challenging. The ambiguity in distinguishing between legitimate and malicious software may explain issues associated with TOOL class. Improved expert annotations could mitigate these issues, enhancing model performance and reliability.

Conversely, we observed standout performance among entities such as THREAT_ACTOR, VULNERABILITY, LOCATION, and MALWARE. Their distinctive names and structured nature (e.g., vulnerabilities are frequently represented as CVE codes) or unambiguous characteristics facilitate their identification within the text. These findings highlight the importance of entity characteristics in model performance and suggest vectors of improvement for NER tasks in CTI contexts.

*2) Comparison on STIXnet Dataset:* In this experiment, we conducted a comparative analysis between the $0\text{-}CTI_{CyBERT}$ model and the STIXnet, a previous state-of-the-art modular system for Information Extraction in CTI. We utilized the dataset provided by [17], which encompasses reports that focus on groups or threat actors from the MITRE ATT&CK framework.

Our approach involved fine-tuning the $0\text{-}CTI_{CyBERT}$ model on the new dataset. The preprocessing steps and training configurations mirrored those of the first experiment.

Table IV presents the comparative results, illustrating that the $0\text{-}CTI_{CyBERT}$ model outperformed the STIXnet model across all classes. This outcome highlights the critical role of context extraction in NLP tasks, such as Entity Extraction in CTI reports. The absence of Transformer-based architectures in the STIXnet approach likely limits its capacity to leverage contextual information effectively.

Moreover, the 0-$CTI_{CyBERT}$ model's adaptability to different entity types and its superior performance on the STIXnet dataset, compared to our original test dataset, underscores the complexity and diversity of the data used for 0-CTI training. The CTI data complexity necessitates a high level of generalization, indicating that models incorporating advanced contextual understanding are better equipped to handle the nuanced demands of Cyber Threat Intelligence.

### B. LLM-as-a-Judge Evaluation of Zero-Shot Systems

Large Language Models (LLMs), such as GPT-4, have shown strong capabilities in understanding and generating unstructured natural language texts [27]. The LLM-as-a-Judge method [28] leverages these models to evaluate the performance of other AI systems, providing qualitative assessments based on criteria such as accuracy, relevance, and coherence.

In our framework, we employed the LLM-as-a-Judge method using ChatGPT-4 to evaluate our zero-shot systems: the zero-shot NER and the zero-shot Relation Extractor. These systems were tested on their ability to accurately extract entities and relations from text without prior annotations.

We evaluated 120 CTI reports from our dataset to ensure a comprehensive assessment. The zero-shot NER system achieved an impressive average score of 0.91 with a standard deviation (STD) of 0.06, indicating high performance and low variability; this suggests that the GLiNER model consistently identifies entities accurately. The zero-shot Relation Extraction module obtained an average score of 0.83 with a higher STD of 0.15, reflecting the inherent complexity of the Relation Extraction task. The number of potential relations within the text increases exponentially with each extracted entity. Additionally, the system's performance is influenced by error propagation from EE to RE due to its modular structure. When an entity is misclassified by the NER, whether supervised or not, it is subsequently passed to the Relation Extraction module, resulting in inevitable errors.

The LLM-as-a-Judge method has proven invaluable for automating the qualitative evaluation, providing insightful assessments that validate the robustness of our zero-shot NER system and identify areas for improvement in the zero-shot Relation Extraction module.

## VI. CONCLUSION AND FUTURE WORK

0-CTI stands out as a unique and powerful tool valuable in enhancing preparedness and resilience within Cybersecurity, offering a range of significant advantages.

Firstly, it automates the extraction of critical information from CTI data, effectively *reducing the burden on security teams and saving them time and effort*. By streamlining this process, 0-CTI ensures that the latest threat intelligence is readily accessible, thereby empowering quicker and more informed decision-making in response to emerging threats.

Central to its appeal are *scalability and modular design features*, which allow for *data-independence* and seamless integration or substitution of various submodules. This adaptability ensures that the system can evolve alongside the dynamic landscape of Cybersecurity, even in the absence of specific expertise or data.

The IOC-finder component enhances 0-CTI's capabilities by swiftly detecting Indicators of Compromise, providing crucial early warnings for potential cyberattacks. Then, the integration of advanced Natural Language Processing (NLP) techniques and models for Entity Extraction and Relation Extraction underscores our solution's transformative role in analyzing and interpreting cyber threats. By leveraging cutting-edge Transformer-based architectures, the system excels in the NER task, *surpassing the performance of alternative approaches* designed for cyber NER challenges. Furthermore, our newly implemented algorithm for Relation Extraction, evaluated qualitatively using a novel LLM-as-a-Judge approach, shows promising results.

By adhering to the Structured Threat Information Expression (STIX) standard, 0-CTI ensures that *extracted knowledge can be efficiently shared* across the cybersecurity community. Its use of a cyber graph format facilitates seamless integration with knowledge bases and dedicated software, supporting *continuous learning* and enhancing system capabilities over time.

In essence, 0-CTI represents a comprehensive and forward-thinking solution that not only enhances operational efficiency but also sets new benchmarks in CTI processing, positioning itself at the forefront of cybersecurity technology.

The 0-CTI system holds potential for advancing and enhancing its capabilities, solidifying its position as a leading solution for CTI Information Extraction. Ongoing experimentation efforts focus on refining the performance with the objective to optimize the system's overall utility. A key area of emphasis addresses a comprehensive evaluation of the system's capabilities in Relation Extraction, involving the acquisition of annotated data to quantify performance evaluation and benchmark against supervised approaches and state-of-the-art methods.

The next milestone involves integrating real-time data sources, enabling more immediate and proactive threat identification. This expansion will extend the system's scope to include diverse and evolving data streams, ranging from social media conversations to activities on the dark web. Each of these sources contributes unique insights into the cyber threat landscape. For instance, social media platforms and news outlets can serve as early indicators of emerging threats or the dissemination of malware campaigns. This evolutionary progression aims to empower the system from a reactive stance, where threats are addressed post-identification, to a proactive approach. Such a transition is crucial in a landscape where early detection plays a pivotal role in minimizing the impact of cyberattacks.